
\magnification=\magstep1
\font\eighteenrm = cmr10 scaled\magstep3
\font\eighteeni = cmmi10 scaled\magstep3
\font\eighteensy = cmsy10 scaled\magstep3
\font\eighteenit = cmti10 scaled\magstep3
\font\eighteenb = cmbx10 scaled\magstep3

\font\twelverm = cmr12

\font\twelvei = cmmi12
\font\twelveit = cmti12
\font\twelveb = cmbx12
\font\twelvesy = cmsy10 scaled\magstep1
\font\twelves = cmsl12

\font\tenrm = cmr10
\font\tenit = cmti10
\font\teni = cmmi10                  
\font\tensy = cmsy10                
\font\tens = cmsl10                  
\font\tenb = cmbx10                  

\font\teniu = cmu10
\font\ninerm = cmr9
\font\ninesy = cmsy9
\font\nineb = cmbx9

\font\eightrm = cmr8
\font\eighti = cmmi8
\font\eightsy = cmsy8

\font\sixrm = cmr6
\font\sixi  = cmmi6
\font\sixsy = cmsy6

\font\fivesy = cmsy5

\def\tenpoint{\def\rm{\fam0\tenrm}
\textfont0=\tenrm \scriptfont0=\eightrm \scriptscriptfont0=\sixrm
\textfont1=\teni \scriptfont1=\eighti \scriptscriptfont1=\sixi
\textfont2=\tensy \scriptfont2=\eightsy \scriptscriptfont2=\sixsy
\textfont3=\tenex \scriptfont3=\tenex \scriptscriptfont3=\tenex
\def\sy{\fam4\tensy}%
\textfont4=\tensy%
\def\sl{\fam5\tens}%
\textfont5=\tens%
\def\bf{\fam6\tenb}%
\textfont6=\tenb%
\def\it{\fam7\tenit}%
\textfont7=\tenit%
\def\prfnt{\fam8\ninesy}%
\textfont8=\ninesy%
\def\hbfnt{\fam9\fivesy}%
\textfont9=\fivesy%
\textfont11=\sixrm \scriptfont11=\sixrm \scriptscriptfont11=\sixrm%
\baselineskip 12pt                                                
\lineskip 1pt
\parskip 5pt plus 1pt
\abovedisplayskip 12pt plus 3pt minus 9pt
\belowdisplayshortskip 7pt plus 3pt minus 4pt \tenrm}
\def\prfnt{\ninesy }
\def\hbfnt{\fivesy }

\def\bigfnt{\twelverm}

\def\eighteenpoint{\def\rm{\fam0\eighteenrm}%
\textfont0=\eighteenrm \scriptfont0=\twelverm \scriptscriptfont0=\eightrm
\textfont1=\eighteeni \scriptfont1=\twelvei \scriptscriptfont1=\eighti
\textfont2=\eighteensy \scriptfont2=\twelvesy \scriptscriptfont2=\eightsy
\textfont3=\tenex \scriptfont3=\tenex \scriptscriptfont3=\tenex
\def\sy{\fam4\eighteensy}%
\textfont4=\eighteensy%
\def\bf{\fam6\eighteenb}%
\textfont6=\eighteenb%
\def\it{\fam7\eighteenit}%
\textfont7=\eighteenit%
\baselineskip 21pt
\lineskip 1pt
\parskip 5pt plus 1pt
\abovedisplayskip 15pt plus 5pt minus 10pt
\belowdisplayskip 15pt plus 5pt minus 10pt
\abovedisplayshortskip 13pt plus 8pt
\belowdisplayshortskip 10pt plus 5pt minus 5pt
\eighteenrm}
\def\twelvepoint{\def\rm{\fam0\twelverm}%
\textfont0=\twelverm \scriptfont0=\tenrm \scriptscriptfont0=\eightrm
\textfont1=\twelvei \scriptfont1=\teni \scriptscriptfont1=\eighti
\textfont2=\twelvesy \scriptfont2=\tensy \scriptscriptfont2=\eightsy
\textfont3=\tenex \scriptfont3=\tenex \scriptscriptfont3=\tenex
\def\sy{\fam4\twelvesy}%
\textfont4=\twelvesy%
\def\sl{\fam5\twelves}%
\textfont5=\twelves%
\def\bf{\fam6\twelveb}%
\textfont6=\twelveb%
\def\it{\fam7\twelveit}%
\textfont7=\twelveit%
\def\prfnt{\fam8\ninesy}%
\textfont8=\ninesy%
\def\hbfnt{\fam9\fivesy}%
\textfont9=\fivesy%
\def\paperfont{\fam10\twelverm}%
\def\dotfont{\fam11\sixrm}%
\textfont11=\sixrm \scriptfont11=\sixrm \scriptscriptfont11=\sixrm%
\baselineskip 15pt
\lineskip 1pt
\parskip 5pt plus 1pt
\abovedisplayskip 15pt plus 5pt minus 10pt
\belowdisplayskip 15pt plus 5pt minus 10pt
\abovedisplayshortskip 13pt plus 8pt
\belowdisplayshortskip 10pt plus 5pt minus 5pt \twelverm}

\tenpoint
\hsize 5.9truein
\vsize 9.5truein
\global \topskip .7truein
\newcount\chapnum
\chapnum = 0
\newcount\sectnum
\sectnum = 0
\newcount\subsecnum            
\subsecnum = 0        
\newcount\eqnum
\newcount\chapnum
\chapnum = 0
\newcount\sectnum
\sectnum = 0
\newcount\subsecnum            
\subsecnum = 0        
\newcount\eqnum
\eqnum = 0
\newcount\refnum
\refnum = 0
\newcount\tabnum
\tabnum = 0
\newcount\fignum
\fignum = 0
\newcount\footnum
\footnum = 1
\newcount\pointnum
\pointnum = 0
\newcount\subpointnum
\subpointnum = 96
\newcount\subsubpointnum
\subsubpointnum = -1
\newcount\letnum
\letnum = 0
\newbox\referens
\newbox\figures
\newbox\tables
\newbox\tempa
\newbox\tempb
\newbox\tempc
\newbox\tempd
\newbox\tempe
\hbadness=10000
\newbox\refsize
\setbox\refsize\hbox to\hsize{ }
\hbadness=1000
\newskip\refbetweenskip
\refbetweenskip = 5pt
\def\ctrline#1{\line{\hss#1\hss}}
\def\rjustline#1{\line{\hss#1}}
\def\ljustline#1{\line{#1\hss}}

\def\ctr#1{\hfill{#1}\hfill}

\def\spose#1{\hbox to 0pt{#1\hss}}
\def\chskipt{\vskip .125in plus 0pt minus 0pt }              
\def\chskipl{\vskip .7in plus 18pt minus 10pt}               
\def\secskipt{\penalty-500\vskip 24pt plus 2pt minus 2pt}    
\def\secskipl{\vskip 3.5pt plus 1pt }
\def\subsecskip{\penalty-500\vskip 6pt plus 2pt minus 2pt }
\def\unchskip{\vskip -.7in }                                 
\def\conskip{\vskip 14pt }
\newif\ifoddeven
\gdef\oneside{\oddevenfalse}

\oneside
\newif\ifnonumpageone

\gdef\nonumberfirst{\nonumpageonetrue}
\nonumberfirst
\output{\ifoddeven\leftright\else\samemarg\fi
	\ifnonumpageone\checkpage\else\empty\fi
	\plainoutput}
\def\leftright{\ifodd\count0{\global\hoffset=\oddmargin}
		       \else{\global\hoffset=\evenmargin}\fi}
\def\samemarg{\global\hoffset=\oddmargin}
\gdef\oddmargin{.25truein}
\gdef\evenmargin{0truein}
\def\checkpage{\ifnum\count0=1\nopagenumbers\else\empty\fi}
\footline={{\pagefont\hss--\hquad\folio\hquad--\hss}}
\def\pagefont{\teniu}
\setbox\referens\vbox{\ctrline{\bf References }\chskipt }
\setbox\figures\vbox{\ctrline{\bf Figure Captions }\chskipt }
\setbox\tables\vbox{\ctrline{\bf Table Captions }\chskipt }

\def\title#1{\ctrline {\titfnt #1} }
\def\titfnt{\eighteenpoint}
\def\author#1{\ctrline{\autfnt #1}\par}
\def\autfnt{\bigfnt}

\def\abstract{
\ctrline{\bf ABSTRACT}\chskipt}

\def\reset{\global\sectnum = 0 \global\eqnum = 0
     \global\subsecnum = 0}
\def\chap#1{\global\advance\chapnum by 1 \reset 
\endpage
\chskipt
\ifodd\count0{
\rightline{{\chapnumfont Chapter \the\chapnum}}
\medskip
\rightline{{\chapfont #1}}}
\else{
\leftline{{\chapnumfont Chapter \the\chapnum}}
\medskip
\ljustline{{\chapfont #1}}}\fi
\penalty 100000 \chskipl  \penalty 100000
{\let\number=0\edef\next{
\write2{\bigskip\noindent
  \tofcfont Chapter \the\chapnum.{ }#1
  \leadtofc\number\count0\smallskip}}
\next}}
\def\titcon#1{\unchskip
\ifodd\count0{
\rightline{{\chapfont #1}}}
\else{
\leftline{{\chapfont #1}}}\fi
\penalty 100000 \chskipl \penalty 100000 }
\def\chapnumfont{\tenit}
\def\chapfont{\eighteenpoint}
\def\chapter#1{\chap{#1}}
\def\sect#1{\global\advance\sectnum by 1 \global\subsecnum = 0
\secskipt
\ifnum\chapnum=0{{\sectfont\par\noindent
\secsign\the\sectnum{ }{ }#1}\par
	 {\let\number=0\edef\next{
	 \write2{\vskip 0pt\noindent\hangindent 30pt
	 \tofcfont\hbox to 30pt{\hfill\the\sectnum\quad}\unskip#1
	 \leadtofc\number\count0}}
	 \next}}
\else{{\sectfont\par\noindent
\secsign\the\chapnum .\the\sectnum{ }{ }#1}\par
      {\let\number=0\edef\next{
       \write2{\vskip 0pt\noindent\hangindent 30pt
       \tofcfont\hbox to 30pt{\hfill\the\chapnum
	      .\the\sectnum\quad}\unskip#1
	 \leadtofc\number\count0}}
       \next}}\fi
       \nobreak\medskip\nobreak}
\def\sectfont{\twelvepoint}
\def\secsign{\S}

\def\subsect#1{\global\advance\subsecnum by 1 \secskipt
\noindent
\ifnum\chapnum=0{
     {\subsecfont\the\sectnum .\the\subsecnum{ }{ }#1}
      \nobreak\par
     {\let\number=0\edef\next{
     \write2{\vskip 0pt\noindent\hangindent 58pt\tofcfont
     \hbox to 60pt{\hfill\the\sectnum
	   .\the\subsecnum\quad}\unskip#1
     \leadtofc\number\count0}}\next}}
\else{{
     \subsecfont\the\chapnum .\the\sectnum
	.\the\subsecnum{ }{ }#1}\nobreak\par
     {\let\number=0\edef\next{
     \write2{\vskip 0pt\noindent\hangindent 58pt\tofcfont
     \hbox to 60pt{\hfill\the\chapnum.\the\sectnum
	   .\the\subsecnum\quad}\unskip#1
     \leadtofc\number\count0}}\next}}\fi
     \nobreak\medskip\nobreak}

\def\subsecfont{\twelvepoint}        
\immediate\openout 2 tofc
\def\tableofcontents#1{\endpage
\count0=#1
\chskipt
\ifodd0{
\rjustline{{\chapfont Contents}}}
\else{
\ljustline{{\chapfont Contents}}}\fi
\chskipl
\rjustline{{\tofcfont Page}}
\bigskip
\immediate\closeout 2
\input tofc
\endpage}

\def\tofcfont{\ninerm}
\def\leadtofc{\leaders\hbox to 8pt{\hfill.\hfill}\hfill}
\immediate\openout 4 refc
\def\refbegin#1#2{\unskip\global\advance\refnum by1
\xdef\rnum{\the\refnum}
\xdef#1{\the\refnum}
\xdef\rtemp{$^{\rnum}$}
\unskip
\immediate\write4{\vskip 5pt\par\noindent\tofcfont
  \hangindent .11\wd\refsize \hbox to .11\wd\refsize{\hfill
  \the\refnum . \quad } \unskip}\unskip
  \immediate\write4{#2}\unskip}

\def\ref#1{\refbegin{\?}{#1}}

\def\refsbegin#1#2{\unskip\global\advance\refnum by 1
\xdef\refb{\the\refnum}
\xdef#1{\the\refnum}
\xdef\rrnum{\the\refnum}
\unskip
\immediate\write4{\vskip 5pt\par\noindent\tofcfont
  \hangindent .11\wd\refsize \hbox to .11\wd\refsize{\hfill
  \the\refnum . \quad } \unskip}\unskip
  \immediate\write4{#2}\unskip}
\def\REFSCON#1#2{\unskip \global\advance\refnum by 1
\xdef#1{\the\refnum}
\xdef\rrrnum{\the\refnum}
\unskip
\immediate\write4{\vskip 5pt\par\noindent\tofcfont
  \hangindent .11\wd\refsize \hbox to .11\wd\refsize{\hfill
  \the\refnum . \quad } \unskip}\unskip
  \immediate\write4{#2}\unskip}

\def\refsend{\nobreak$^{\refb-\the\refnum}$\unskip}

\def\foot#1{\footnote{$^{\the\footnum}$}{#1}
  \global\advance\footnum by 1}

\def\figure#1#2{\global\advance\fignum by 1
\xdef#1{\the\fignum }
\ctrline{\Figure . #2}\par\conskip
{\let\number=0\edef\next{
\write3{\par\noindent\tofcfont
  \hangindent .11\wd\refsize \hbox to .11\wd\refsize{\hfill
  \the\fignum . \quad } \unskip}
\write3{#2\leadtofc\number\count0\par}}
\next}}
\def\figurs#1#2#3{\global\advance\fignum by 1
\xdef#1{\the\fignum }
\ctrline{\Figure . \it #2}
\ctrline{\it #3}\par\conskip
{\let\number=0\edef\next{
\write3{\par\noindent\tofcfont
  \hangindent .11\wd\refsize \hbox to .11\wd\refsize{\hfill
  \the\fignum . \quad}\unskip}
\write3{#2 #3\leadtofc\number\count0\par}}
\next}}

\def\figcon{\ctrline{{\it Figure  \the\fignum} -- cont'd}\par\conskip}
\def\Figure{{\it Figure  \the\fignum}}
\immediate\openout 3 figc

\def\table#1#2{\global\advance\tabnum by 1
\xdef#1{\the\tabnum }
\ctrline{\Table . #2}\par\conskip
{\let\number=0\edef\next{
\write5{\par\noindent\tofcfont
  \hangindent .11\wd\refsize \hbox to .11\wd\refsize{\hfill
  \the\tabnum . \quad } \unskip}
\write5{#2\leadtofc\number\count0\par}}
\next}}
\def\tabls#1#2#3{\global\advance\tabnum by 1
\xdef#1{\the\tabnum }
\ctrline{\Table . #2}
\ctrline{\it #3}\par\conskip
{\let\number=0\edef\next{
\write5{\par\noindent\tofcfont
  \hangindent .11\wd\refsize \hbox to .11\wd\refsize{\hfill
  \the\tabnum . \quad}\unskip}
\write5{#2 #3\leadtofc\number\count0\par}}
\next}}

\def\Table{\it Table  \the\tabnum}
\immediate\openout 5 tabc

\def\eqname#1{\global\advance\eqnum by 1
\ifnum\chapnum=0{
   \xdef#1{ (\the\eqnum ) }(\the\eqnum )  }
\else{
   \xdef#1{ (\the\chapnum .\the\eqnum ) }
	    (\the\chapnum .\the\eqnum ) }\fi}
\def\enum{\global\advance\eqnum by 1
  \ifnum\chapnum=0{ (\the\eqnum )  }
  \else{(\the\chapnum .\the\eqnum ) }\fi}

\def\eqn#1{\eqno \eqname{#1} }
\def\eqnameap#1{\global\advance\eqnum by 1
   \xdef#1{ (\copy\appbox .\the\eqnum ) }
	    (\copy\appbox .\the\eqnum ) }
\def\enumap{\global\advance\eqnum by 1
  (\copy\appbox .\the\eqnum ) }

\def\item#1{\par\noindent\hangindent .08\wd\refsize
\hbox to .08\wd\refsize{\hfill #1\quad}\unskip}
\def\sitem#1{\par \noindent\hangindent .13\wd\refsize
\hbox to .13\wd\refsize{\hfill #1\quad}\unskip}
\def\ssitem#1{\par\noindent\hangindent .195\wd\refsize
\hbox to .195\wd\refsize{\hfill #1\quad}\unskip}

\def\point{\par \global\advance\pointnum by 1
\noindent\hangindent .08\wd\refsize \hbox to .08\wd\refsize{\hfill
\the\pointnum .\quad}\unskip}

\def\spoint{\par \global\advance\subpointnum by 1
\noindent\hangindent .13\wd\refsize
\hbox to .13\wd\refsize{\hfill
(\char\the\subpointnum )\quad}\unskip}

\def\sspoint{\par \global\advance\subsubpointnum by 1
\noindent\hangindent .195\wd\refsize
\hbox to .195\wd\refsize {\hfill\hbox to 20pt{
(\romannumeral\subsubpointnum\hfill)}\quad}\unskip}

\def\bye{\endpage\end}              

\def\bspace#1{\hbox to -#1{}}
\newbox\appbox
\def\appendix#1{\endpage\reset
\setbox\appbox\hbox{#1}
\chskipt \ctrline {\bf APPENDIX #1 }\penalty 10000
\chskipl \penalty 10000
\write2{\bigskip\noindent
 {\tofcfont Appendix #1\leadtofc\number\count0\par\smallskip}}}

\def\mat#1#2{\if 2#1 {\left( \  \vcenter{\halign{$\ctr{## }$ \quad
& $\ctr{## }$\cr #2}} \  \right) } \else{ }\fi
\if 3#1 {\left( \  \vcenter{\halign{
$\ctr{## }$ \quad & $\ctr{## }$ \quad
& $\ctr{## }$\cr #2}} \  \right) } \else{ }\fi
\if 4#1 {\left( \  \vcenter{\halign{$\ctr{## }$ \quad &
$\ctr{## }$ \quad & $\ctr{## }$ \quad
& $\ctr{## }$\cr #2}} \  \right) } \else{ }\fi
\if 5#1 {\left( \  \vcenter{\halign{$\ctr{## }$ \quad
& $\ctr{## }$ \quad & $\ctr{## }$ \quad
& $\ctr{## }$ \quad & $\ctr{## }$\cr #2}} \  \right)} \else{ }\fi
\if 6#1 {\left( \  \vcenter{\halign{$\ctr{## }$ \quad
& $\ctr{## }$ \quad & $\ctr{## }$ \quad & $\ctr{## }$ \quad
& $\ctr{## }$ \quad & $\ctr{## }$\cr #2}} \  \right)} \else{ }\fi }

\def\endpage{\par \vfill \eject}
\def\physrev{\baselineskip 24pt
\lineskip 1pt
\parskip 1pt plus 1pt
\abovedisplayskip 15pt plus 7pt minus 13.33pt
\belowdisplayskip 15pt plus 7pt minus 13.33pt
\abovedisplayshortskip 14pt plus 11pt
\belowdisplayshortskip 9pt plus 7pt minus 7pt
\def\chskipt{\vskip 24pt }
\def\chskipl{\vskip 6.5pt }
\def\secskipt{\vskip 7pt plus 3pt minus 1.33pt }
\def\secskipl{\vskip 3.5pt plus 2pt }
\def\subsecskip{\vskip 7pt plus 2pt minus 2pt }
\def\unchskip{\vskip -6.5pt }
\def\conskip{\vskip 24pt }
\refbetweenskip = \the\baselineskip
\multskip\refbetweenskip by 5
\divskip\refbetweenskip by 10
\twelverm }
\def\bk{\hfil\break}         

\def\To{\par\noindent\hangindent .18\wd\refsize
\hbox to .18\wd\refsize {To: \hfill \qquad }}
\def\from{\par\noindent\hangindent .18\wd\refsize
\hbox to .18\wd\refsize {From: \hfill \qquad }}
\def\topic{\par\noindent\hangindent .18\wd\refsize
\hbox to .18\wd\refsize{Topic: \hfill \qquad }}

\def\startpage#1 {\count0 = #1}
\def\startchapter#1{\chapnum = #1 \advance\chapnum by -1}

\def\startfig#1{\fignum = #1 \advance\fignum by -1}
\def\starttab#1{\tabnum = #1 \advance\tabnum by -1}

\newbox\xa\newbox\xb
\def\boxit#1{ \setbox\xa\vbox {\vskip \boxitsep
\hbox{\hskip \boxitsep #1\hskip \boxitsep }\vskip \boxitsep }
\setbox\xb\hbox{\vrule \copy\xa \vrule}
\vbox{\hrule width 1\wd\xb \copy\xb \hrule width 1\wd\xb }}
\def\fboxit#1#2{ \setbox\xa\vbox {\vskip \boxitsep
\hbox{\hskip \boxitsep #2\hskip \boxitsep }\vskip \boxitsep }
\setbox\xb\hbox{\vrule width #1pt \copy\xa \vrule width #1pt}
\vbox{\hrule height #1pt width 1\wd\xb
\copy\xb \hrule height #1pt width 1\wd\xb }}
\def\reboxit#1#2#3{ \setbox\xa\vbox{\vskip \boxitsep
\hbox{\hskip \boxitsep #3\hskip \boxitsep }\vskip \boxitsep }
\setbox\xb\hbox{\vrule width #1pt\bspace{#2}
\copy\xa \vrule width #1pt}
\vbox{\hrule height #1pt width 1\wd\xb
\copy\xb \hrule height #1pt width 1\wd\xb}}
\def\boxitsep{4pt}

\newdimen\offdimen
\def\offset#1#2{\offdimen #1
   \noindent \hangindent \offdimen
   \hbox to \offdimen{#2\hfil}\ignorespaces}
\newdimen\defnamespace   
\defnamespace=2in        
\def\definition#1#2{     
    \def\itema{\par\hang\textindent}
    {\advance\parindent by \defnamespace
     \advance\defnamespace by -.5em
     \itema{\hbox to \defnamespace{#1\hfil}}#2\par}}
\def\TEX{\hbox{T\hskip-2pt\lower1.94pt\hbox{E}\hskip-2pt X}}
\def\wyz{\hbox{WI\hskip-1pt\lower.9pt\hbox{Z\hskip-1.85pt
\raise1.7pt\hbox{Z}}LE}}
\def\\{$\backslash $}

\def\underwiggle#1{\mathop{\vtop{\ialign{##\crcr
    $\hfil\displaystyle{#1}\hfil$\crcr\noalign{\kern2pt\nointerlineskip}
    $\scriptscriptstyle\sim$\crcr\noalign{\kern2pt}}}}\limits}
\def\({[}
\def\){]}

\def\hquad{\hskip.5em{}}
\mathchardef\app"3218

\def\linebreak{\break}
\mathchardef\oprod="220A
\mathchardef\inter="225C
\mathchardef\union="225B

\mathchardef\relv="326A
\mathchardef\leftv="326A
\mathchardef\rightv="326A
\mathchardef\relvv"326B
\mathchardef\leftvv"326B
\mathchardef\rightvv"326B
\mathchardef\Zscr"25A
\mathchardef\Yscr"259
\mathchardef\Xscr"258
\mathchardef\Wscr"257
\mathchardef\Vscr"256
\mathchardef\Uscr"255
\mathchardef\Tscr"254
\mathchardef\Sscr"253
\mathchardef\Rscr"252
\mathchardef\Qscr"251
\mathchardef\Pscr"250
\mathchardef\Oscr"24F
\mathchardef\Nscr"24E
\mathchardef\Mscr"24D
\mathchardef\Lscr"24C
\mathchardef\Kscr"24B
\mathchardef\Jscr"24A
\mathchardef\Iscr"249
\mathchardef\Hscr"248
\mathchardef\Gscr"247
\mathchardef\Fscr"246
\mathchardef\Escr"245
\mathchardef\Dscr"244
\mathchardef\Cscr"243
\mathchardef\Bscr"242
\mathchardef\Ascr"241
\mathchardef\lscr"160

\immediate\openout 2 tofc
\immediate\openout 3 figc
\immediate\openout 4 refc
\immediate\openout 5 tabc
\tolerance 4000
\def\refbegin#1#2{\unskip\global\advance\refnum by1
\xdef\rnum{\the\refnum}
\xdef#1{\the\refnum}
F\xdef\rtemp{[\rnum]}
\unskip
\immediate\write4{\vskip 5pt\par\noindent\tofcfont
  \hangindent .11\wd\refsize \hbox to .11\wd\refsize{\hfill
  \the\refnum . \quad } \unskip}\unskip
  \immediate\write4{#2}\unskip}
\def\refsend{\nobreak[\refb-\the\refnum]\unskip}
\tolerance 4000
\def\refbegin#1#2{\unskip\global\advance\refnum by1
\xdef\rnum{\the\refnum}
\xdef#1{\the\refnum}
\xdef\rtemp{[\rnum]}
\unskip
\immediate\write4{\vskip 5pt\par\noindent\tofcfont
  \hangindent .11\wd\refsize \hbox to .11\wd\refsize{\hfill
  \the\refnum . \quad } \unskip}\unskip
  \immediate\write4{#2}\unskip}
\def\refsend{\nobreak[\refb-\the\refnum]\unskip}

\def\rar{\rightarrow}
\def\_#1{_{\scriptscriptstyle #1}}
\def\^#1{^{\scriptscriptstyle #1}}
\def\fk{f\_k(\vec r,\vec v,t)}
\def\els{\ell\_*}
\def\cmss{cm/s\^{2}}
\def\kms{km/s}
\def\vsm{\langle v\^2\rangle}
\def\FF{F}
\def\nineb{\bf}
\def\ninerm{}
\def\({[}
\def\){]}
\def\msun{M_\odot}
\def\lsun{L_\odot}
\def\dtr{d\^3r}
\def\dtv{d\^3v}
\def\div{\vec\nabla\cdot}
\def\grad{\vec\nabla}

\def\h0{H_{o}}
\def\ao{a_{o}}
\def\vin{V\_{\infty}}
\def\vpar{v\_{\parallel}}
\vsize 23.6truecm
\hsize 5.6truein
\parskip 0pt
\def\oddmargin{.3in}
\def\evenmargin{.1in}
\def\ML{M_*}
\def\vh{v\_{h}}
\def\bk{\par\noindent}
\vskip 0.5truein
\centerline{{\bf MODIFIED-DYNAMICS PREDICTIONS AGREE WITH}}
\centerline{{\bf OBSERVATIONS OF THE
HI KINEMATICS IN FAINT DWARF GALAXIES}}
\centerline{{\bf CONTRARY TO THE CONCLUSIONS OF LO SARGENT AND YOUNG}}
\vskip 0.5truein
\centerline{ Mordehai Milgrom}
\bk
\centerline{Department of Physics, Weizmann Institute of Science}
\centerline{ 76100 Rehovot, Israel}
\centerline{and}
\centerline{D\'epartment d'Astrophysique Relativiste et de Cosmologie}
\centerline{Observatoire de Paris, 92195 Meudon, France }
\vskip 0.5truein
\vskip 30pt
\baselineskip 10pt
\ninerm
{\bf \noindent Abstract.}
 Lo, Sargent, and Young (1993) have recently published an analysis of the
HI kinematics of nine faint dwarf galaxies.
Among other things, they conclude that the masses of these systems, as
 deduced by the modified dynamics (MOND) from the observed
velocity dispersions,
are systematically smaller
 than even the HI masses that are observed in these
systems, by a factor of ten or more.
 Such a state of things would speak strongly against MOND.
We show here that the MOND mass estimator used by Lo et al.
is smaller
than the proper expression, by a factor of about
 twenty.
We derive the proper mass estimator as an exact virial-like relation
 between the  3-D rms velocity, $\vsm$, and the total mass, $M$, of an
arbitrary, self-gravitating system,
made of light constituents, that is everywhere in the
very-low-acceleration regime of MOND. This reads
$M=(9/4)\vsm\^2/G\ao$. (For a system that is not stationary, $\vsm$ involves
 also an average over time.)
We do this in the
Bekenstein-Milgrom formulation of MOND as a modification of gravity.
This relation has been known before
for the special case of a stationary, spherical
system. We further generalize this relation to cases with
constituent masses
that are not small compared with that of the whole system.
We discuss
various applications of the $M-v$ relation; inter alia, we derive an
expression for
the two-body force law in the large-distance limit.
With the correct estimator
 the predictions of MOND are, by and large,
 in  good agreement with the
total observed masses (the observed gas mass plus a stellar mass
corresponding to an M/L of order one solar unit).

\vskip 20pt
{\nineb I. Introduction}
\vskip 5pt
\par
Lo, Sargent, and Young (1993) (hereafter LSY) have recently published the
 results of HI observations of nine intrinsically faint dwarf galaxies.
Among other things, they determine total HI masses. They also
estimate and discuss the total masses implied by the modified dynamics (MOND).
They find that the latter are systematically  smaller than the HI masses ``by
a factor of ten or more''--an unacceptable state of affairs
(for the theory).
\par
LSY have used a MOND mass estimator that is a mistaken adaptation of
the relation
$$M=\vin^4/G\ao,  \eqn{\i}$$
between the total mass, $M$, of a body,
 and the asymptotic rotational
velocity of test particles around it, $\vin$ (Milgrom 1983a,b);
 $\ao$ is the acceleration constant of the theory.
\par
Equation \i is exact in MOND, but LSY
have used an analogous relation with $\vin$
replaced by
$\vh$, the half width at half maximum of the integrated HI
line profile--which is a measure of the {\it one-dimensional} rms velocity
for the whole galaxy. LSY thus use
$$\ML=\vh\^4/G\ao, \eqn{\ii} $$

We show below that $\ML$ gives a far underestimate of the mass.
We derive a general mass-velocity relation
that holds in the non-relativistic, Bekenstein-Milgrom (1984)
 formulation of MOND as a
modification of gravity; this relation is
$$M={9 \over 4}\vsm\^2/G\ao.   \eqn{\iii}$$
Here, $\vsm$ is the {\it three-dimensional}
mean-square velocity of the whole system
(averaged over time if the system is not stationary),
and $M$ is its total mass.
 Relation \iii applies for an arbitrary, self-gravitating
system, whereby the acceleration
 is much smaller than $\ao$ everywhere. Beside the factor of 9/4
 missing in expression \ii,
 a large factor of disparity (9 in the isotropic case)
 is present because of the difference between
 3-D and 1-D velocities, leading to an undrestimate of the masss by a factor
 of about twenty, in the isotropic case.
\par
Gerhard and Spergel (1992) have derived relation \iii for the special
case of a stationary, spherical system; it was
derived, for the yet more special case
of a sphere with constant
radial and tangential velocity dispersions, by Milgrom (1984).
\vskip 20pt
{\nineb II. The  MOND mass estimator}
\vskip 5pt
\par
Consider a self-gravitating system that is composed of
various particle species with masses $m_k$ and
distribution functions $\fk$.
As usual, take the time derivative of the quantity
$$Q\equiv\sum_k \int\dtr~\dtv~m_k~\fk~\vec r\cdot\vec v \eqn{\coo} $$
(itself, the time derivative of the trace of the moment-of-inertia tensor).
$$\dot Q=\sum_k \int\dtr~\dtv~m_k~\fk~v\^2+
\sum_k \int\dtr~\dtv~m_k~\fk~\vec r\cdot\vec a. \eqn{\x} $$
(By Liouville's theorem the time derivative of a quantity of the form
$\int\dtr\dtv fq(\vec r,\vec v,t)$ is $\int\dtr\dtv f\dot q$.)
The first term in eq.\x is the momentary,
mass-weighted, 3-D, mean-square velocity, $\vsm$,
multiplied by the total mass, $M$. In the second term we put
 $\vec a=-\grad\phi$ (here it is assumed that gravity is the only important
 force),
 where $\phi$ is the (MOND) gravitational potential,
to obtain
$$\dot Q=M\vsm(t)-
\sum_k \int\dtr~\dtv~m_k~\fk~\vec r\cdot\grad\phi. \eqn{\xi} $$
As $\vec a=-\grad\phi$ depends only on $\vec r$,
the $v$ integration, and the sum over species can now be performed
 to give the standard result
$$\dot Q=M\vsm(t)-
 \int\dtr~\rho(\vec r,t)\vec r\cdot\grad\phi. \eqn{\xii} $$
We now specialize to MOND, and here enters
the assumption of self gravity: the density $\rho$ is
the source of the gravitational potential. We substitute $\rho(\vec r,t)$
 from the field equation
of Bekenstein and Milgrom (1984),
$$\div\(\mu(x)\grad\phi\)=4\pi G\rho,
 ~~~~x\equiv\vert\grad\phi\vert/\ao,  \eqn{\xiii}$$
in eq.\xii; then, integrating by parts, we get
$$\dot Q=M\vsm(t)-{1\over 4\pi G}\int\_{\Sigma}\mu(x)
 \vec r\cdot\grad\phi~\grad\phi\cdot d\vec s
+{1\over 4\pi G}
\int\mu(x)\grad\phi\cdot\grad(\vec r\cdot\grad\phi)~\dtr. \eqn{\xv}$$
The first integral is over any surface, $\Sigma$, encompassing all the mass;
we take it at infinity. As $r$ goes to infinity $\grad\phi$
becomes $(MG\ao)\^{1/2}\vec r/r\^2$ (see Bekenstein and Milgrom 1984),
and $\mu(x)$ becomes $x$, so the second
term becomes $-M(MG\ao)\^{1/2}$. The second integral
can be calculated thus
$$I_2\equiv
\int\mu(x)\grad\phi\cdot\grad(\vec r\cdot\grad\phi)~\dtr=
\int\mu(x)\grad\phi\cdot\((\vec r\cdot\vec \nabla)\grad\phi+
(\grad\phi\cdot\vec\nabla)\vec r\)~\dtr=$$
$$\int{1\over 2}\mu(x)\vec r\cdot\vec\nabla
\((\grad\phi)\^2\)+\mu(x)(\grad\phi)\^2~\dtr.  \eqn{\xvi}$$
Let $\FF(y)$ be such that
 $\mu(x)=\FF'(y)\vert{}_{y=x^2}~$
\($\FF(x\^2)$ is the Lagrangian density for the
potential (see Bekenstein and Milgrom 1984)\).
The first term in the integrand of eq.\xvi can then be written as
$${\ao\^2\over 2}\vec r\cdot\vec\nabla\FF=
{\ao\^2\over 2}\div(\FF\vec r)-{3\ao\^2\over 2}\FF.  \eqn{\xvii} $$
The first divergence term may again be written as a surface integral
at infinity that contributes
$(1/3)M(MG\ao)\^{1/2}$ to the right-hand side of eq.\xv.
We thus end up with the relation
$$\dot Q/M=\vsm(t)-{2\over 3}(MG\ao)\^{1/2}-
{\ao\^2\over 4\pi GM}\int{3\over 2}\FF(x\^2)-\mu(x)x\^2~\dtr
 \eqn{\xviii} $$
 (with $x\equiv\vert\grad\phi\vert/\ao$).
 Take now the long-time average of eq.\xviii.
 That of the left-hand side vanishes,
 as $Q$ is finite at both ends of time,
 and we get
 $$\overline{\vsm}={2\over 3}(MG\ao)\^{1/2}+
{\ao\^2\over 4\pi GM}\int{3\over 2}\overline{\FF(x\^2)}
-\overline{\mu(x)x\^2}~\dtr, \eqn{\xxi} $$
 where an overline signifies the long-time average.
This relation is exact and general (i.e independent on how close
to the MOND regime we are).
In the Newtonian limit ($\ao$ goes to 0) it gives the usual virial theorem.
In the very limit where
 the accelerations are always and everywhere
 in the system much smaller than $\ao$,
the integrands in eqs. \xviii \xxi vanish \(because
in the limit $x\rar0$ we have $\FF(x\^2)\rar(2/3)x\^3$, and
$\mu(x)\rar x$\),
 and we obtain
 $$\overline{\vsm}={2\over 3}(MG\ao)\^{1/2}.  \eqn{\xx} $$
 When the system is stationary (and sometimes in more general cases--such
 as systems that are stationary in some rotating frame)
 the momentary value of $\vsm$ remains constant
 and it can then be used in eq.\xx. As we cannot determine time averages for
 relevant astrophysical systems, we
 usually make the simplifying assumption that
 the system is stationary and use relation \xx with the observed momentary
 value of $\vsm$.
\par
Equation\xx is the trace of the following relation , which holds under
the same conditions, and which can be
derived along the same lines:
$$\overline{\langle v\_i v\_j\rangle}={2\over 9}(MG\ao)\^{1/2}\delta\_{ij}
+{\ao\^2\over 8\pi GM}\int \overline{(\delta\_{ij}-
3\phi,\_i\phi,\_j/\vert\grad\phi\vert\^2)F}~\dtr. \eqn{\bg}$$
\par
There are possible formulations of MOND other than that of
Bekenstein and Milgrom (1984).
The present $M-v$ relation is probably not exact in all
of them. It is, however, a good indicative estimator, and, in any case,
the best we have at the moment. We note, for example, that in the class
of MOND theories based on modification of the law of inertia, discussed
by Milgrom (1993), the $M-v$ relation is still exact for stationary,
spherical systems, whose constituents move on circular trajectories
(in the deep MOND limit).
\par
In many instances, what we would like to use as test particles are
 themselves sub-systems
with internal structure, or with masses that are not small compared with $M$.
We must then be careful
in treating the subsystems.
Let $m$ be the mass of a body we want to treat as a
structureless constituent, and let $\ell$ be its size. The assumptions
underlying our derivation of the $M-v$ relation seem to require the
following:
(i) The constituent itself is kept together by gravitational forces balancing
internal motions (say of some perfect fluid).
(ii) In order to have $\vert\grad\phi\vert\ll\ao$ everywhere (within
the body as well as outside) we must have $mG/\ell\^2\ll\ao$.
(iii) If $L$ is the length scale over which the field varies appreciably,
in the vicinity of the body, we must have $\ell\ll L$.
(iv) As we want to use only the centre-of-mass velocity of the body in
calculating $\vsm$, and neglect the internal velocities, we must have
$m\ll M$.
\par
This set of requirements is too restrictive for the $M-v$ relation to be of
much application: Condition ii, for instance,
would bar stars from being legitimate
constituents; condition i would bar atoms and elementary particles, etc.
(condition iii is quite benign).
Fortunately, it is possible to weaken conditions i,ii greatly, or, indeed,
to eliminate them altogether. Also, the $M-v$ relation may be generalized
to circumvent condition iv.
\par
We note first that when all the would-be constituents have $m\ll M$
 requirements i,ii may be disregarded. To see this, replace all bodies that
do not satisfy
i and/or ii by ones that do; i.e., by ones having the same mass, but
a size $\els$ satisfying $mG/\els\^2\ll\ao$, and that are made of a perfect
fluid held by gravity. Because $m\ll M$, $\els$ can be chosen small enough
so that $\els\ll L$. The new system does satisfy all the assumptions,
and the $M-v$ relation holds for it. However, If $m$ is small enough
there exist a length $a$ such that $a\ll L$ on one hand, and
such that at a distance $a$ from $m$ the contribution of the
latter to $\grad\phi$ is small. Under these conditions, Bekenstein and
Milgrom (1984) showed that the centre-of-mass motion of the body is oblivient
to
its internal structure and mass, and it behaves like a test particle. The
centre-of-mass motions of our replacement bodies are, then, like that
of the original. The correction on this very-small-mass case
is of order $(m/M)\^{1/2}$. In addition, we want to be able to neglect the
internal velocities in the replacement constituent bodies; again this
is permitted because $m\ll M$, and the higher order correction is here
also of order $(m/M)\^{1/2}$. It is thus valid to apply the
$M-v$ relation to the original system without having to worry about
the internal makeup of the constituents.
\par
When at least some of the sub-system's masses are not very small compared
with $M$, the $M-v$ relation, as given by eq.\xx is not valid.
Given the masses and sizes
of the constituents, we consider the limit where the
distances between them are very large--in keeping with our working in the
extreme MOND limit. We can, as described before, replace the masses
with ``live'', self-gravitating ones made of a perfect fluid.
Relation \xx can now be applied
but we must reckon with the internal velocities of the ``live''
replacements,
which does not necessarily have a counterpart in the original
constituent. In the limit of large distances, each replacement body
itself satisfies eq.\xx, as can be shown. (We assume that the internal
acceleration produced by the constituent dominates over that of the rest,
at a distance where the asymptotic form of the isolated-mass
potential--as described below eq.\xv--already holds.
This is valid in the limit that we consider here.)
The prescription is thus as follows: if
$m\_i$ is the mass of the $i$th particle, and $v\_i$ its centre-of-mass
velocity, then the mean square velocity in the $M-v$ relation is
to be taken as
$$M\^{-1}\sum\_i m\_i (v\_{i}\^{2}+{2\over 3}\sqrt{m\_i G\ao}), \eqn
{\xxiii}$$
where the second term in parentheses is the internal rms velocity
within $m\_i$.
This leads to the relation
$${2\over 3}(MG\ao)\^{1/2}(1-M\^{-3/2}\sum m\_{i}\^{3/2})=
M\^{-1}\sum m\_{i}v\_{i}\^{2}=\vsm,  \eqn{\xxiv}$$
where $\vsm$ now includes only centre-of-mass velocities.
\par
To demonstrate the neccesity of this crucial modification,
consider a system made
of two masses $m$ and $M\gg m$, in a circular orbit around each other.
A blind application of relation \xx will give $MG\ao\approx(9/4)(m/M)\^2v\^4$,
where $v$ is the speed of the smaller mass. This, however, is
quite wrong:
MOND tells us that, in fact,
$MG\ao\approx v\^4$. The reason
 for the failure of eq\xx, in this case, is that most of the
contribution to $\vsm$ should have come from intrinsic motions within $M$,
which are neglected when we treat it as a test particle.
If we apply eq.\xxiv instead, we get the correct result (as the first order
term in $m/M$).
\par
The relative correction introduced by eq.\xxiv is bounded by
$(m\_x/M)\^{1/2}$,
where $m\_x$ is the maximum mass of a constituent.
When all the masses are equal the relative correction is $(m/M)\^{1/2}$.
\par
At any rate, when the accelerations internal to the subsystems are
small compared with $\ao$, it is best not to consider them as test particles,
but include in $\vsm$ the full rms velocity for the system.
\vskip 20pt
{\nineb III. Some applications of the mass-velocity relation}
\vskip 5pt
\par
First, and perhaps foremost, the general
$M-v$ relation puts on firmer and wider footings the MOND prediction
of a Faber-Jackson like relation for all self-gravitating
low-acceleration objects. This includes disk galaxies, as we saw; for these,
the $M-v$ relation comes in addition to the
(differnt) relation between total mass
and asymptotic rotational velocity \i.
These relations tell us that we can use
in the Tully-Fisher relation either the
 rms roatational velocity (for low-$a$ galaxies) or the asymptotic, rotational
velocity (for all disks). We then expect relations with the same slope, but
different intercepts. In fact, almost all known
astronomical objects, on the scale of galaxies and up, have
internal, mean accelerations that are about $\ao$ or smaller, and the $M-v$
relation should approximately hold universally. In applying it to observed
objects we note again the uncertainties in the deduction of the 3-D
velocities, and also the error
introduced if we assume that a system is already
stationary when, in fact, they are not.
(and take its momentary value of $\vsm$ to be the time-average
value).
\par
The $M-v$ relations affords significant shortcuts to
the derivation of the MOND forces acting on bodies
in a few configurations.
If the gravitational forces can be balanced by rotating a configuration,
rigidly, with some frequencey $\omega$, about some
known centre, then, $\omega$
may be determined from relation \xxiv, and from it the forces on all
the masses may be gotten.
Perhaps the most interesting application is a derivation of an
 analytic expression for the two-body force in
the Bekenstein-Milgrom formulation, in the long distance limit:
Consider two masses $m\_1$, and $m\_2$, at a distance $\ell$ from each
other: a system that can be supported by rotation about the centre of mass.
In the limit of large $\ell$
an expression for the
force, $F$, can be conveniently put in the form
$$F(m\_1,m\_2,\ell)={m\_1 m\_2\over \ell}
\left(G\ao\over m\_1+m\_2\right)^{1/2}
A(m\_1/m\_2), \eqn{\ci} $$
with
$$A(q)={2\over 3}q\^{-1}(1+q)\^{1/2}
\((1+q)\^{3/2}-q\^{3/2}-1\).  \eqn{\xxv}$$
Numerical results for the dimensionless function
$A(q)$, for a few values of the mass ratio, $q$, are
given by Milgrom (1986), and agree with those determined from the
analytic expression\xxv.
We note that $A(q)$ varies rather little across its full range,
  from $A=1$ at $q=0$
to $A\approx0.8$, at $q=1$. ($A$ is, of course, symmetric
under the exchange of the two masses $q\leftrightarrow q\^{-1}$,
because the forces on the two masses are equal.)
\par
Similarly, we can calculate the force (of a more academic interest) on any
of $n$ equal masses $M/n$, on the vertices of a regular polygon,
of diameter $2r$, with a mass $m$ at its centre.
$$F(M,m,n,r)=
{2\over 3n}(M+m)\^{3/2}
(G\ao)\^{1/2}\left\(1-{m\^{3/2}+n\^{-1/2}M\^{3/2}\over(M+m)\^{3/2}}
\right\)r\^{-1}. \eqn{\xxxv} $$
This was checked numerically, by R. Brada (private communication),
for some configurations.
\par
It is useful to specialize the $M-v$ relation to rotationally supported
disks (e.g. a very low acceleration disk galaxies). If $\Sigma(r)$ is the
surface density, and $v(r)$ the rotation curve, then eq.\xx implies
the relation
$${2\over 3}M\^{3/2}(G\ao)\^{1/2}=\int\_{0}\^{\infty}2\pi r\Sigma(r)
v\^2(r)~dr,  \eqn{\xxx} $$
where $M=\int\_{0}\^{\infty}2\pi r\Sigma(r)~dr$,  is the disk's mass.
If we add a point mass $m$ at the centre, representing a bulge, say
(within a radius that is small
compared with the dimensions of the disk).
eq.\xxiv tells us that the
relation is now
$${2\over 3}\((M+m)\^{3/2}-m\^{3/2}\)(G\ao)\^{1/2}
=\int\_{0}\^{\infty}2\pi r\Sigma(r)v\^2(r)~dr.  \eqn{\xxxi} $$
\par
If we consider a special case of a thin ring of mass $M$ and radius $r$,
with a point mass
$m$ in the centre,
we can use eq.\xxxi to calculate the
gravitational force on the ring ($M$ times the force on a unit mass):
$$F={2\over 3}(M+m)\^{3/2}
(G\ao)\^{1/2}\left\(1-\left({m\over M+m}\right)\^{3/2}
\right\)r\^{-1}.   \eqn{\xxxii} $$
(This is also a special case of eq.\xxxv, taking there the
limit of $nF$ for $n\rightarrow\infty$.)

\vskip 20pt
{\nineb IV. A revistation of the masses of the dwarfs}
\vskip 5pt
\par
 The 3-D rms velocity, which enters the $M-v$ relation
 is not, in itself, directly observable, and
 we have to express $\vsm$ in terms of the observed, integrated, 1-D
 rms velocity, $\vpar$. This, as usual, involves assumptions on the
 velocity structure of the system.
 If $\vpar$ is independent of the direction
 of the line of sight--as when the velocity
distribution is isotropic--we have $\vsm=3\vpar\^2$, and the
mass estimator for this case is
$$M={81 \over 4}(\vpar\^4/G\ao). \eqn{\iv}$$
\par
If the motions in the system are confined to a plane that is at an
inclination $i$ to the line of sight, and if, in that plane, the
overall velocity distribution is isotropic--as is the case for any
axisymmetric systems, such as a rotation-supported disk--then
$\vsm=(2/sin^2~i)\vpar\^2$, and
$$M= {9\over sin^4~i}(\vpar\^4/G\ao).    \eqn{\vi} $$
We see that the LSY estimator could, in the case of flat
systems, be even more
 discrepant with the correct one. For example for $i<45^o$ we get
 a factor of discrepancy
$>36$ (for $i<30^o$ the factor is $>144$).
\par
Two factors conjoin to give the large disparity between
the naive mass estimator used by LSY, and the correct expression.
The estimator given in eq\i, which LSY mimic,
 uses the asymptotic circular velocity around
the system. However,
the (3-D) rms velocity in a system is smaller then the
asymptotic {\it rotational} velocity corresponding to system's mass.
Also, the measured line-of-sight rms velocity is of course smaller
then the 3-D one. Since these ratios enter
in the fourth power, the resulting
disparity is large.
 As $\vpar$ is approximately
$\vh$ we see that the mass estimator used by LSY differs by a large
factor from the correct one.
\(In fact, $\vpar$ may be somewhat smaller than $\vh$--according to LSY
 $\vh\approx 1.18\vpar$--but the correction introduced
by this difference is compensated by the fact that LSY use the outdated
pristine estimate of $\ao$ given by Milgrom 1983b, when the best value
today, which is based on detailed studies of rotation curves,
 is about a factor of two smaller (Begeman, Broeils, and Sanders 1991)\).
\par
In default of more detailed information on the galaxies studied by LSY, the
best we can do to estimate their MOND masses is to use relation \iv.
This means that we simply have to multiply LSY's MOND masses by a factor
of twenty.
Because deduced MOND masses are subject to large errors, as they depend on
the fourth power of the velocities, we feel it is more appropriate
to use the ``luminous'' masses in relation \iv in order to calculate
 from them the velocity dispersions
that MOND implies, and compare these with the observed line-of-sight
velocities $\vpar$ (designated $\sigma\_T$ by LSY).
We do this in Table 1 where we give
$\vpar\equiv\sigma\_T$ along with the total line-of-sight
velocity dispersions predicted by MOND on the basis of the stationary
$M-\vpar$ relation \iv, for the isotropic case.
These are calculated for three values
of the stellar $M/L$: $\alpha=0.5,~1.,~4$ solar units,
and are designated $\vpar(\alpha)$.(We multiply
the HI mass by a factor of 1.3 to account for He.)
We use the value
$\ao=1.2~10\^{-8}\cmss$ (Begeman, Broeils, and Sanders 1991).
We see that the observed $\vpar$ falls within the predicted range
of velocities corresponding to the range of $M/L$ values we use,
in all but two
galaxies, where MOND predict higher velocities than observed (the quoted
errors in $\vpar$ are about $\pm20\%$).
\par
We must note that the comparison for individual cases is subject to
large uncertainties due to the following factors:
(i) We do not know that the systems under study are in a stationary state,
so that their momentary, observed rms velocity equals the long-time average.
In fact, LSY state that the velocity fields of some of the galaxies suggest
radial motions such as expansion or contraction.
Non-stationary systems spend more time in a state of
lower-than-average rms velocity. If departure from stationarity is important
we would expect that the velocities deduced from the $M-v$
relation--representing long-time averages--would be, by and large, larger
than the momentary, observed ones.
(ii) Our $M-v$ realtion assumes that the HI is the dominant mass component
(i.e. is self gravitating).
This is clearly not so in some of the cases (for which LSY find
$M\_H/L\_B<1$); notably in LGS-3, and DDO216, but also in DDO69, and DDO155).
In these cases, the deduced velocities depend strongly on the distributions
of both the stellar, and the HI masses which we do not know.
Here there is an additional uncertainty in the value of the
observed masses due to that in the stellar $M/L$, which we try to cover
in Table 1.
(iii) The distances to the galaxies in the sample are poorly
known according to LSY. The value of the deduced velocities is proportional
to the square root of the observed distance.
(iv) The observed line-of-sight velocities need not be $3\^{-1/2}$ of the
3-D rms velocities as we assume in applying the $M-\vpar$ relation for
the isotropic case.

\vskip 5pt
{\bf Acknowledgement} The impetus to prove a general $M$-$v$ relation
stemed from the fact that numerical calculations of the proportionality
factor in the $M$-$v$ relation,
for different disk models, gave the same value as for the spherical
case. I thank Rafael Brada for performing the numerical calculations.
The hospitality of the department of relativistic astrophysics and cosmology
at the Paris Observatory is gratefully acknowledged.

\vskip 20pt
\noindent
{\bf References}
\vskip 4pt
\bk
Begeman, K.G., Broeils, A.H., and Sanders, R.H., 1991,
MNRAS, {\nineb 249}, 523
\bk
Bekenstein, J., and Milgrom, M., 1984, ApJ {\nineb 286}, 7
\bk
Gerhard, O.E., and Spergel, D.N. 1992, ApJ {\nineb 397}, 38
\bk
Lo, K.Y., Sargent, W.L.W., and Young, K. 1993, AJ {\nineb 106}, 507
\bk
Milgrom, M., 1983a, ApJ {\nineb 270}, 365
\bk
Milgrom, M., 1983b, ApJ {\nineb 270}, 371
\bk
Milgrom, M., 1984, ApJ {\nineb 287}, 571
\bk
Milgrom, M., 1986, ApJ {\nineb 302}, 617
\bk
Milgrom, M., 1993, Annals of Physics, In the press
\endpage
$$\vbox{\settabs
\+&UGC4483-----&-----------&-----------&---------------&-----------&-----------&\cr 
\+&Name & $M\_{HI}$ & $L\_B$ & $\vpar$ & $\vpar(.5)$
&$\vpar(1)$&$\vpar(4)$\cr
\vskip .5truecm
\+&LGS-3   & 0.02 & 0.07 & $7.5\pm 1.3$  & 4.7 & 5.3 & 7.0 &          \cr
\+&UGC4483 & 3.7  & 2.3  & 13.0 & 14.5& 15.4& 18.2&          \cr
\+&DDO69   & 3.6  & 3.6  & 8.0  & 15.0& 16.0& 19.7&          \cr
\+&CVn~dwA & 8.1  & 1.1  & 9.5  & 17.2& 17.4& 18.5&          \cr
\+&DDO155  & 0.2  & 0.23 & 10.5 & 7.4 & 7.9 & 9.9 &          \cr
\+&DDO187  & 5.0  & 2.3  & 13.0 & 15.6& 16.2& 18.7&          \cr
\+&Sag~D1G & 0.8  & 0.25 & 8.0  & 7.8 & 10.1& 11.3&          \cr
\+&DDO210  & 0.3  & 0.13 & 8.0  & 7.7 & 8.0 & 9.2 &          \cr
\+&DDO216  & 1.3  & 6.9  & 10.0 & 14.2& 16.1& 21.9&          \cr
      }$$
\par
Table 1. Observed masses, luminosities, and
line-of-sight velocity dispersions
for the dwarfs, along with velocity dispersions predicted by MOND
$\vpar(\alpha)$, for assumed $M/L=\alpha$ in solar units.
 All masses in units of $10\^7~\msun$, luminosities in units of
$10\^7~\lsun$, and velocities in $\kms$.
\bye